\documentclass[aps,twocolumn,superscriptaddress]
{revtex4-1}
\usepackage{amsfonts}
\usepackage{dcolumn}
\usepackage{upgreek}
\usepackage{bm}
\usepackage{tikz}
\usepackage{ulem}
\usepackage[colorlinks=true,citecolor=blue,urlcolor=blue]{hyperref}
\usepackage{amsmath,amsfonts,amssymb,times}
\usepackage{natbib} 
\usepackage{jabbrv}
\usepackage[standard]{ntheorem}

\makeatletter
\newcommand*{\Rom}[1]{\expandafter\@slowromancap\romannumeral #1@}
\makeatother

\begin{document}

\title{Realization of edge states along a synthetic orbital angular momentum dimension}


\author{Yu-Wei Liao}

\affiliation{CAS Key Laboratory of Quantum Information, University of Science and Technology of China, Hefei 230026, China}
\affiliation{CAS Center For Excellence in Quantum Information and Quantum Physics, University of Science and Technology of China, Hefei 230026, China}

\author{Mu Yang} \email{myang@ustc.edu.cn}
\affiliation{CAS Key Laboratory of Quantum Information, University of Science and Technology of China, Hefei 230026, China}
\affiliation{CAS Center For Excellence in Quantum Information and Quantum Physics, University of Science and Technology of China, Hefei 230026, China}

\author{Hao-Qing Zhang}
\affiliation{CAS Key Laboratory of Quantum Information, University of Science and Technology of China, Hefei 230026, China}
\affiliation{CAS Center For Excellence in Quantum Information and Quantum Physics, University of Science and Technology of China, Hefei 230026, China}

\author{Zhi-He Hao}
\affiliation{CAS Key Laboratory of Quantum Information, University of Science and Technology of China, Hefei 230026, China}
\affiliation{CAS Center For Excellence in Quantum Information and Quantum Physics, University of Science and Technology of China, Hefei 230026, China}

\author{Jun Hu}
\affiliation{CAS Key Laboratory of Quantum Information, University of Science and Technology of China, Hefei 230026, China}
\affiliation{CAS Center For Excellence in Quantum Information and Quantum Physics, University of Science and Technology of China, Hefei 230026, China}

\author{Tian-Xiang Zhu}
\affiliation{CAS Key Laboratory of Quantum Information, University of Science and Technology of China, Hefei 230026, China}
\affiliation{CAS Center For Excellence in Quantum Information and Quantum Physics, University of Science and Technology of China, Hefei 230026, China}

\author{Zong-Quan Zhou}
\affiliation{CAS Key Laboratory of Quantum Information, University of Science and Technology of China, Hefei 230026, China}
\affiliation{CAS Center For Excellence in Quantum Information and Quantum Physics, University of Science and Technology of China, Hefei 230026, China}
\affiliation{Hefei National Laboratory, University of Science and Technology of China, Hefei 230088, China}

\author{Xi-Wang Luo}
\affiliation{CAS Key Laboratory of Quantum Information, University of Science and Technology of China, Hefei 230026, China}
\affiliation{CAS Center For Excellence in Quantum Information and Quantum Physics, University of Science and Technology of China, Hefei 230026, China}

\author{Jin-Shi Xu}\email{jsxu@ustc.edu.cn}
\affiliation{CAS Key Laboratory of Quantum Information, University of Science and Technology of China, Hefei 230026, China}
\affiliation{CAS Center For Excellence in Quantum Information and Quantum Physics, University of Science and Technology of China, Hefei 230026, China}
\affiliation{Hefei National Laboratory, University of Science and Technology of China, Hefei 230088, China}

\author{Chuan-Feng Li}\email{cfli@ustc.edu.cn}
\affiliation{CAS Key Laboratory of Quantum Information, University of Science and Technology of China, Hefei 230026, China}
\affiliation{CAS Center For Excellence in Quantum Information and Quantum Physics, University of Science and Technology of China, Hefei 230026, China}
\affiliation{Hefei National Laboratory, University of Science and Technology of China, Hefei 230088, China}

\author{Guang-Can Guo}
\affiliation{CAS Key Laboratory of Quantum Information, University of Science and Technology of China, Hefei 230026, China}
\affiliation{CAS Center For Excellence in Quantum Information and Quantum Physics, University of Science and Technology of China, Hefei 230026, China}
\affiliation{Hefei National Laboratory, University of Science and Technology of China, Hefei 230088, China}

\date{\today}

\begin{abstract}
The synthetic dimension is a rising method to study topological physics, which enables us to implement high-dimensional physics in low-dimensional geometries. 
Photonic orbital angular momentum (OAM), a degree of freedom characterized by discrete yet unbounded, serves as a suitable synthetic dimension.
However, a sharp boundary along a synthetic OAM dimension has not been demonstrated, dramatically limiting the investigation of topological edge effects in an open boundary lattice system. In this work, we make a sharp boundary along a Floquet Su-Schrieffer-Heeger OAM lattice and form approximate semi-infinite lattices by drilling a pinhole on the optical elements in a cavity. The band structures with zero ($\pm\pi$) energy boundary states are measured directly, benefiting from the spectra detection of the cavity. Moreover, we obtain the edge modes moving from the gap to the bulk by dynamically changing the boundary phase, and we reveal that interference near the surface leads to spectrum discretization. Our work provides a new perspective to observe edge effects and explore practical photonics tools.
\end{abstract}

\maketitle
\textit{Introduction---}Synthetic dimensions are newly developed tools to study topological materials in recent years. The central idea of the synthetic dimension is to exploit a set of physical states based on particles' internal degrees of freedom to simulate the motion along an extra lattice. Synthetic dimensions have been formed based on degrees of freedom such as spin~\cite{mancini2015observation,boada2012quantum}, energy levels~\cite{stuhl2015visualizing} of atoms, frequency~\cite{yuan2016photonic,dutt2019experimental,dutt2020single}, orbital angular momentum (OAM)~\cite{yang2022topological,yang2023realization}, arrival times~\cite{chalabi2019synthetic,weidemann2022topological,regensburger2012parity, leefmans2022topological}, and transverse spatial supermodes~\cite{lustig2019photonic} of photons. 
The synthetic dimensions are independent of accurate geometric dimensions, which enables the study of higher-dimensional physics in a lower-dimensional system. Moreover, from flexible modulation and abundant detection methods, plenty of unique topological phenomena have been obtained along synthetic dimensions in atomic and photonic systems~\cite{ozawa2019topological,yuan2018synthetic,yang2022simulating}.

The OAM states of light is one of the approaches to implementing synthetic dimensions, and the first photonics proposal for a topological model with a synthetic dimension was based on the OAM states in a cavity~\cite{luo2015quantum}. A large number of theoretical schemes have been proposed to explore topological physics and devices in the synthetic OAM dimension, such as the Hofstadter butterfly spectrum~\cite{celi2014synthetic}, Weyl semimetal phase~\cite{sun2017weyl}, high-order filters~\cite{luo2017synthetic}, and OAM optical switches~\cite{luo2018topological}. Later on, the synthetic OAM dimensions were experimentally demonstrated and used to explore some bulk topological features, including Zak phase~\cite{cardano2017detection}, and band structures~\cite{yang2022topological}.

For topological materials~\cite{wang2023proximity}, the non-trivial bulk topology can usually be characterized by topological invariants, such as the Chern number~\cite{hatsugai1993chern} or $Z_{2}$ index~\cite{kane2005z}. The edge excited states appear at the contact boundaries of materials with different topological invariants, called the bulk-edge correspondence~\cite{thouless1982quantized}. These edge states are topologically protected and can be used to realize optoelectronic devices with distinct characteristics such as immunity to local defects and high transmission efficiency~\cite{dai2022topologically}. Thus, a sharp boundary is of central importance to study topological phenomena. To date, experimentally creating boundaries along synthetic OAM dimensions remains challenging. 

In this study, we experimentally realize a Su-Schrieffer-Heeger (SSH)-like model and implement a boundary along the synthetic OAM dimensions in a cavity. Compared with topological edge states realized in the atomic synthetic dimension~\cite{mancini2015observation,stuhl2015visualizing,kanungo2022realizing}, photonic frequency~\cite{dutt2022creating} synthetic dimensions and spatial supermodes~\cite{lustig2019photonic} synthetic dimensions, we form a unique approximate semi-infinite lattice and exploring corresponding spectrum properties, which provides a new perspective to attain topological edge effects in SSH model. Forming an approximate semi-infinite lattice along the synthetic dimension poses challenges, which relies on breaking a particular coupling between certain physical states in an adequately long chain. To achieve this, we construct a boundary by drilling a circular hole in the optical element, which cuts off the polarization coupling at the zero OAM mode. This proposal is inspired by the theoretical way to construct a sharp boundary by employing hollow beam splitters~\cite{zhou2017dynamically}. 
We detect the phase transition of edge state energies directly benefiting from the resonant spectrum detection of the cavity, which is only illustrated as theoretical graphs in previous relevant works~\cite{groning2018engineering, cardano2017detection,le2020topological}. Moreover, we obtain the dynamic behaviors of the edge states entrance to the bulk by changing the
phase of the surface and find that interference near a surface leads to the discretization of
the detected energy spectrum, which has not been directly detected in previous experiments.

\begin{figure}[t]
\centering\includegraphics[width=8cm]{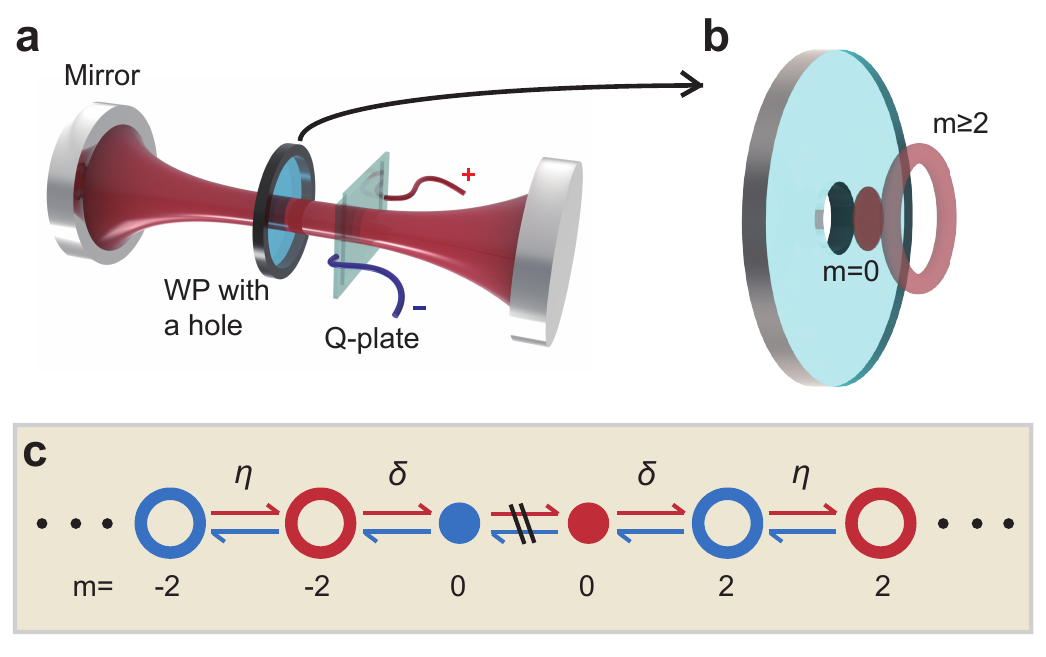}
\caption{The degenerate cavity designed to form the OAM synthetic lattice and the corresponding theoretical model. \textbf{a}. The degenerate cavity contains diverse OAM modes. A Q-plate (q=1) and a wave plate are settled for coupling polarized OAM modes. WP: wave plate. \textbf{b}. The WP with a centered hole in the cavity, where only optical modes with topological charge $|m|\geq 2$ pass through the WP and their polarization can be coupled. \textbf{c}. The schematic of the synthetic lattice formed by OAM modes with different spins (red for left circularly polarized modes, and blue for right circularly polarized modes); $\eta$ and $\delta$ are coupling strengths of the wave plate and Q-plate, respectively. The 1-D chain in synthetic space is cut off between the different polarized fundamental Gaussian modes ($m=0$). 
}
\label{concept}
\end{figure}

\begin{figure*}[t]
\centering\includegraphics[width=18cm]{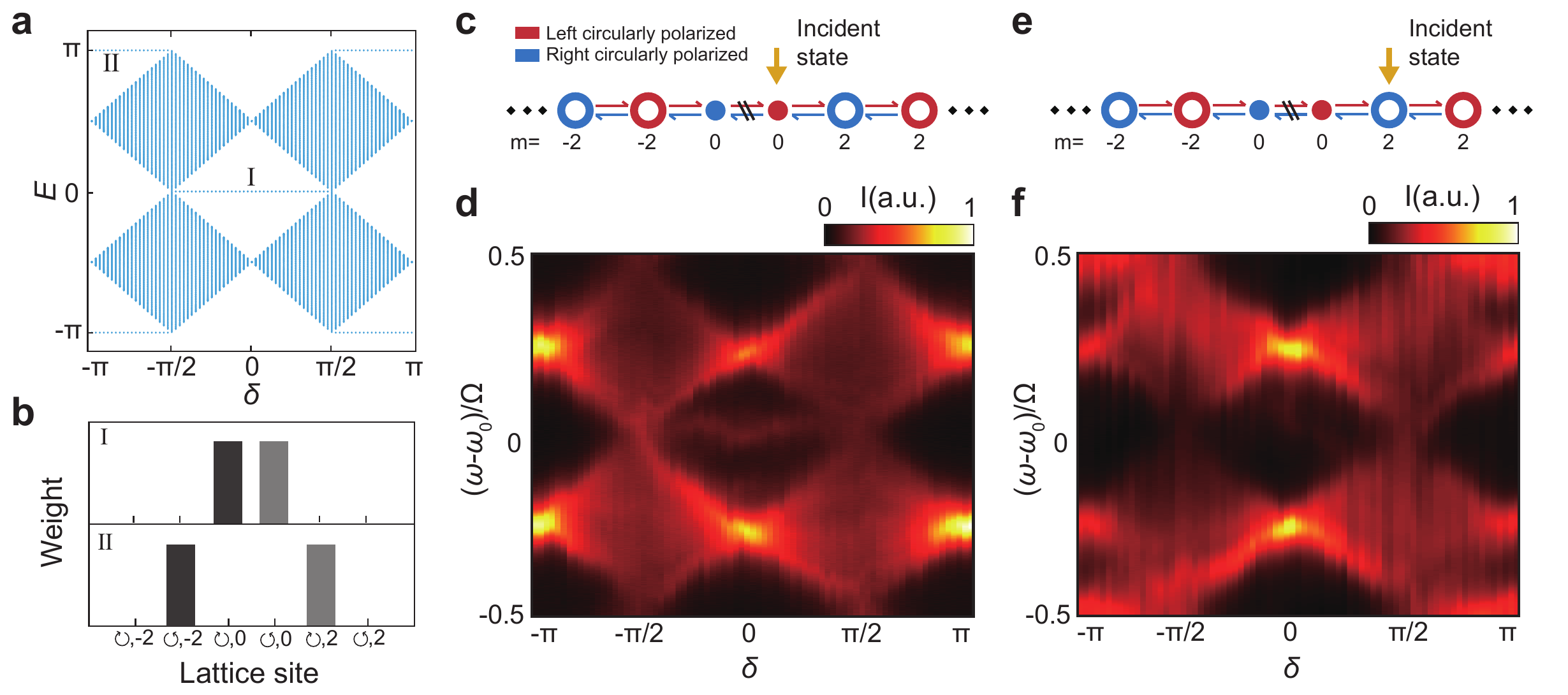}
\caption{Experimental observation of the energy band structures of the edge states. 
\textbf{a}. The theoretical energy spectrum with edge states (\Rom{1} and \Rom{2}). 
\textbf{b}. Distributions of the edge states with 0 (\Rom{1}) or $\pm\pi$ (\Rom{2}) energy when $\delta=0$ or $\pm\pi$.
\textbf{c, e}. The schemes of the OAM lattice model when exciting different sites.  
\textbf{d, f}. The normalized transmission intensity spectra with edge state \Rom{1} or \Rom{2} when pumping the cavity with a left circularly polarized fundamental Gaussian mode (d) or right circularly polarized $m=2$ OAM mode (f). $\omega-\omega_{0}$ is the frequency detuning, and $\Omega=375$ MHz represents the free spectral range (FSR).
}
\label{spectrum}
\end{figure*}

\begin{figure}[t]
\centering\includegraphics[width=8cm]{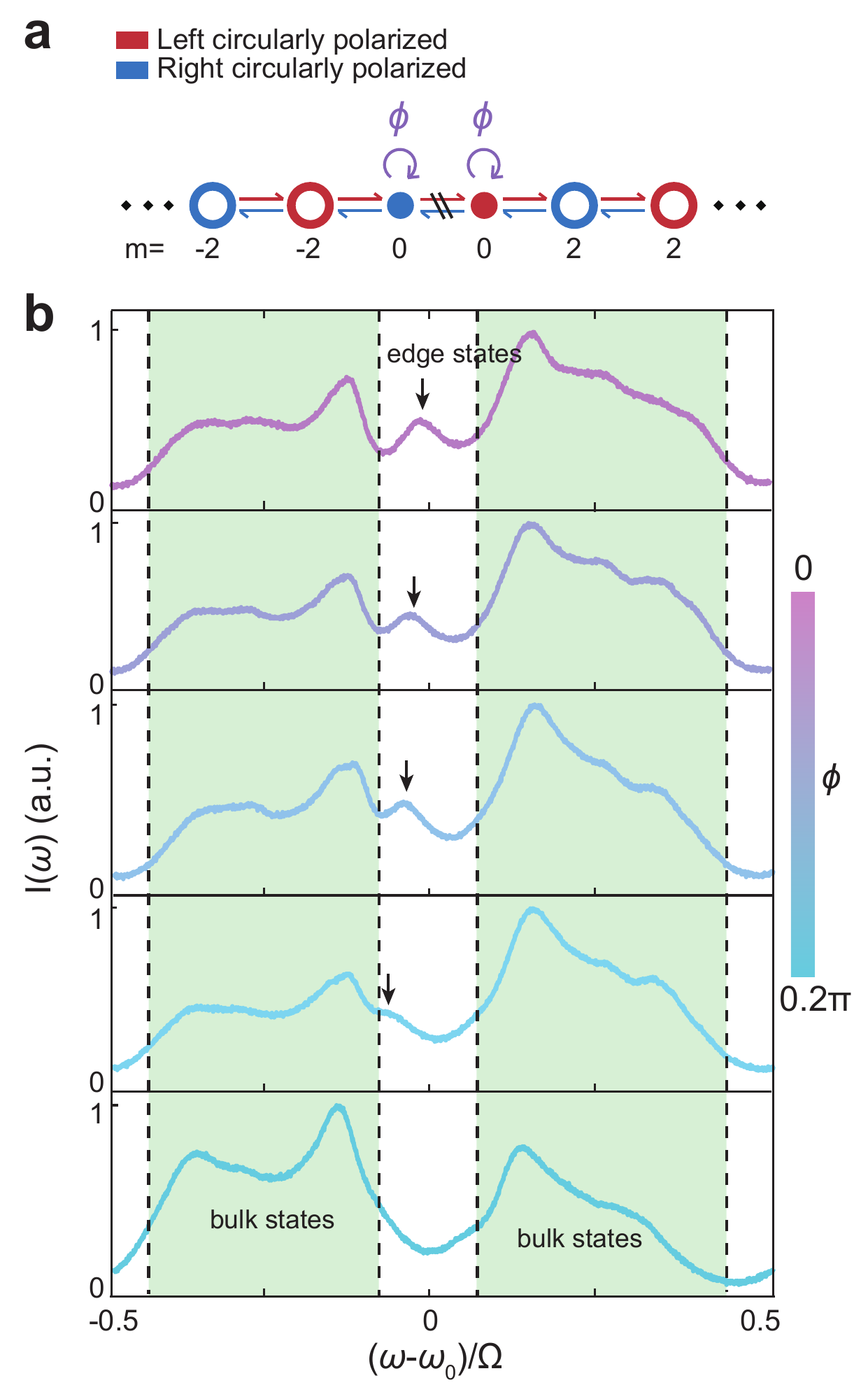}
\caption{The experimentally measured edge state behaviors with boundary perturbations. 
\textbf{a}. The schemes of the OAM lattice model when modulating the phase $\phi$ of the states on the boundary ($m=0$).  \textbf{b}. The detected transmission spectra with different $\phi$ when $\delta=-\pi/3$. The edge state (indicated by the arrow) moves from $\omega=\omega_0$ into bulk states as $\phi$ varies from 0 to 0.2$\pi$. $\omega-\omega_{0}$ is the frequency detuning, and $\Omega=375$ MHz represents the free spectral range (FSR).   
}
\label{move}
\end{figure}

\textit{Creating boundaries in a synthetic OAM lattice---} 
We start our experiment in a degenerate optical cavity (Fig. \ref{concept} \textbf{a}) \cite{cheng2017degenerate,cheng2018experimental,Cheng:19}, which can support a set of OAM modes with different topological charges $m$ to resonate at the same frequency (see supplementary materials (SM) \Rom{5} for details). To form a one-dimensional (1-D) chain in OAM synthetic space, we introduce a wave plate (WP) made of birefringent materials and a Q-plate consisting of anisotropic liquid crystal molecules into the cavity, respectively, as shown in Fig. \ref{concept} \textbf{a}. Compared with the platform based on a stack of Q-plates and wave plates~\cite{cardano2017detection}, the introduction of the cavity enables the reuse of optical elements which greatly saves experimental resources.
The optical OAM modes with different circular polarizations [$\circlearrowleft$ (left circular polarization) or $\circlearrowright$ (right circular polarization)] are coupled with each other by the WP with coupling strength $\sin(\eta/2)$, where $\eta$ is the phase delay between the ordinary and extraordinary light. On the other hand, OAM modes with adjacent even topological charges $m$ and $m+2$ are coupled for the spin-orbital interaction of the Q-plate with coupling strength $\sin(\delta/2)$, where $\delta$ is an electrically controlled parameter. The coupled OAM modes form a 1-D SSH-like lattice, as shown in Fig. \ref{concept} \textbf{c}. 

Our objective is to cut off the connection between two specific adjacent modes, thereby establishing a boundary within the synthetic lattice system. The OAM modes with $m=0$ are fundamental Gaussian modes with the intensity peaks distributed in the center. In contrast, the OAM modes with $m\neq0$ have doughnut-shape intensity distributions, where the intensities peak located on a ring with the radius scaled as $\sqrt{m}$ (see SM section \Rom{6} for the intensity profiles of the different cavity transverse OAM modes). Given the spatial distinguishability of the intensity distributions between the $m=0$ and $m\neq0$ modes, we can implement an m-dependent operation within the cavity to truncate the OAM chain.

Here, we make a hole at the center of the WP, as shown in Fig. \ref{concept}b. The diameter of the hole is specially designed so that the wave plate will only modulate the polarization of photons with $|m|\geq2$. In contrast, photons with $m=0$ can directly go through the hole and remain unchanged. With a suitable pinhole with radius $r=0.75\omega_{0}=125$ $\upmu$m, where $\omega_{0}$ is the radius of the fundamental Gaussian mode, the undesired modal cross-coupling caused by the overlap between the cavity transverse OAM modes and the spatial aperture can be almost ignored. (see SM section \Rom{6} for details).
As illustrated in Fig. \ref{concept}c, the polarization coupling of $m=0$ is destroyed, and a sharp boundary is created. On the other hand, the largest OAM index $m_{\rm max}$ (corresponding to radius $r_{\rm max}=\omega_{0}\sqrt{m_{\rm max}/2}$) is limited by the effective diameter $R=6.35$ mm of the Q-plate, and the second boundary appears on the sites of $m_{\rm max}>5000$. Thus, the OAM lattice can be viewed as two approximate semi-infinite lattices, where the interface between the non-trivial topological bulk and “vacuum” can support edge states. For brevity, the Hamiltonian for this system in tight binding approximation ($\delta, \eta\ll 1$) can be written as (see SM section \Rom{2} for more details)
\begin{equation}
\begin{aligned}
\hat{H}=&\sum_{m}\delta(a_{\circlearrowright,m+2}^{\dag}a_{\circlearrowleft,m}+\mathrm{h.c.})\\
&+\sum_{m\neq0}\eta(a_{\circlearrowright,m}^{\dag}a_{\circlearrowleft,m}-a_{\circlearrowleft,m}^{\dag}a_{\circlearrowright,m}),
\end{aligned}\label{H0}
\end{equation}
where the $a_{\circlearrowleft(\circlearrowright),m}$ and $a_{\circlearrowleft(\circlearrowright),m}^{\dag}$ represent the creation and annihilation operators of the $m$-th OAM modes with specific polarization ($\circlearrowleft$ and $\circlearrowright$). As parameters $\delta$ and $\eta$ become larger, the system changes to the Floquet version and the Hamiltonian cannot be written as Eq. \ref{H0}. Instead, we define an effective Hamiltonian as $H_{\rm eff}=-i\ln{(J_{Q}J_{W}J_{W}J_{Q})}$, where $J_{Q}$ and $J_{W}$ are the operators of the Q-plate and WP  (see SM section \Rom{1} for the matrix forms of the operators). As an example, the theoretical eigen-energies $E$ of the effective Hamiltonian $H_{\rm eff}$ with $\delta\in(-\pi,\pi]$ when $\eta=\pi/2$ is shown in Fig. \ref{spectrum} a. In addition to the bulk states, the zero and $\pm\pi$ energy edge states appear in the gap as shown in the position \Rom{1} and \Rom{2}, respectively. 

As $\delta$ changes from $-\pi$ to $\pi$, the energy state positions of the edge modes will change abruptly at the gap closing points ($\delta=\pm\pi/2$), corresponding to the topological phase transition of the system. Aside from the difference in the edge state energies, the located positions of the edge state in real space are also different. As shown in Fig. \ref{spectrum}b, the edge states of zero energy (\Rom{1}, $\delta=0$) are located at $m=0$ site, while the edge states of $\pm\pi$ energy (\Rom{2}, $\delta=\pm\pi$) are located at $m=\pm2$ sites.

\textit{Probing the synthetic lattice with boundaries---} 
Benefiting from the resonant spectrum detection of the cavity, the band structures and density of the states can be directly measured.
The input and output relations satisfy Langevin equations
\begin{equation}
\begin{aligned}
    &\partial_{t}{a_n(t)}=-i[a_n,\hat{H}]-\frac{\gamma}{2}a_n(t)-\sqrt{\gamma}b_{in,n}(t)\\
    &b_{out,n}(t)=\sqrt{\gamma}a_n(t),  
\label{L_equation}
\end{aligned}
\end{equation}
where $n=s,m$ and $s\in\{\circlearrowleft,\circlearrowright\}$ is the polarization index. $a_{n}$ represents the light field of polarized OAM modes in the cavity and the $b_{in(out),n}$ is the in- (out-) put light field. $\gamma$ is the decay rate of each mode in the cavity. Transferring to the frequency domain, the input-output relation can be written as $b_{out,n}(\omega)=\sum_{n^{'}}T_{nn^{'}}b_{in,n^{'}}(\omega)$, where $T_{nn^{'}}=-i\langle n|\gamma/(\omega-\hat{H}+i\gamma/2)|n^{'}\rangle$ represents the transmission coefficient (see SM section \Rom{3} for more details). 
We can pre-select the input state $|n'\rangle$ and reveal the edge state distribution features according to the corresponding transmission spectrum, where the edge state self-energy appears only when the corresponding edge mode is excited.
In the experiment, we pump the cavity with a sweeping continuous-wave laser of left circularly polarized Gaussian modes. The site ($n=\circlearrowleft,0$) on the boundary along the OAM lattice is excited as shown in Fig. \ref{spectrum}c. The corresponding transmitted intensity spectrum $I(\omega)=\sum_{n}|T_{nn^{'}}|^2$ varying with parameter $\delta$ is shown in Fig. \ref{spectrum}d. The system energy corresponds to $E=2\pi(\omega-\omega_{0})/\Omega$, where $\omega-\omega_{0}$ is the frequency detuning, and $\Omega=375$ MHz represents the free spectral range (FSR) of the cavity. It is worth noting that the energy period $(-\pi,\pi]$ of the Floquet system corresponds to the frequency detuning range $(-\Omega/2,\Omega/2]$. The energy spectrum of bulk states agrees well with theoretical predictions in Fig. \ref{spectrum}a. 
Moreover, only the zero energy edge when $\delta\in(-\pi/2, \pi/2)$ can be obtained with pre-selecting state $n=\circlearrowleft,0$, which demonstrates the zero energy edge state mainly distributed on the boundary ($m=0$), as predicted in Fig. \ref{spectrum}a-b (I). As we change the input modes to the right circularly polarized OAM modes $n=\circlearrowright,2$ (Fig. \ref{spectrum}e), only the $\pm\pi$ energy edge modes when $\delta\in(-\pi, -\pi/2)$ or $\delta\in(\pi/2, \pi)$ appear (Fig. \ref{spectrum}f), which demonstrate energy edges with $\pm\pi$ energy mainly distribute on the second-nearest boundary ($|m|=2$), as predicted in Fig. \ref{spectrum}a-b (\Rom{2}). 
On the other hand, we can reveal the distribution of edge states by post-selecting the output state $|n\rangle$ on different bases, denoted as projective measurements (see SM \Rom{7} for details).

\textit{Edge state behaviors with boundary perturbations---} The edge state is the topological feature in the SSH model and is protected by chiral symmetry. Therefore, the edge state will be significantly affected when the symmetry changes. To investigate the robustness of the edge state energy, we probe the band structure with perturbations on the boundary. 

One advantage of photonic synthetic dimensions is that we can dynamically change the lattice structures by flexibly adjusting the optical parameters. 
By tilting the angle of the WP with a central hole, the thickness of the birefringent crystal changes continuously. The light of high-order OAM modes ($m\neq0$) passing through the crystal accumulates a roundtrip phase $e^{-i\phi}$ compared with the fundamental Gaussian mode passing through the hole, which corresponds to an effective phase shift for the $m=0$ unit cell. Thus, phase perturbations on the boundary are introduced into the OAM lattice, as illustrated in Fig. \ref{move} \textbf{a}. 

The experimentally transmitted intensity spectra $I(\omega)$  with $\delta=-\pi/3$ of different $\phi$ are shown in the Fig. \ref{move}b. As $\phi=0$, a small transmitted peak corresponding to the zero energy edge state (marked by the black arrow) is located in the center of the topological band gap. As the parameter $\phi$ changes from 0 to $0.2\pi$, we observe the zero energy edge state move to the bulk states (green regions) and the topological protection is effectively destroyed. This is because the roundtrip phase causes a shift of the quasi-energies for the $m=0$ unit cell, which breaks the chiral symmetry of the model as the dangling sites are at different energies to the bulk modes.

\begin{figure}[t]
\centering\includegraphics[width=8.9cm]{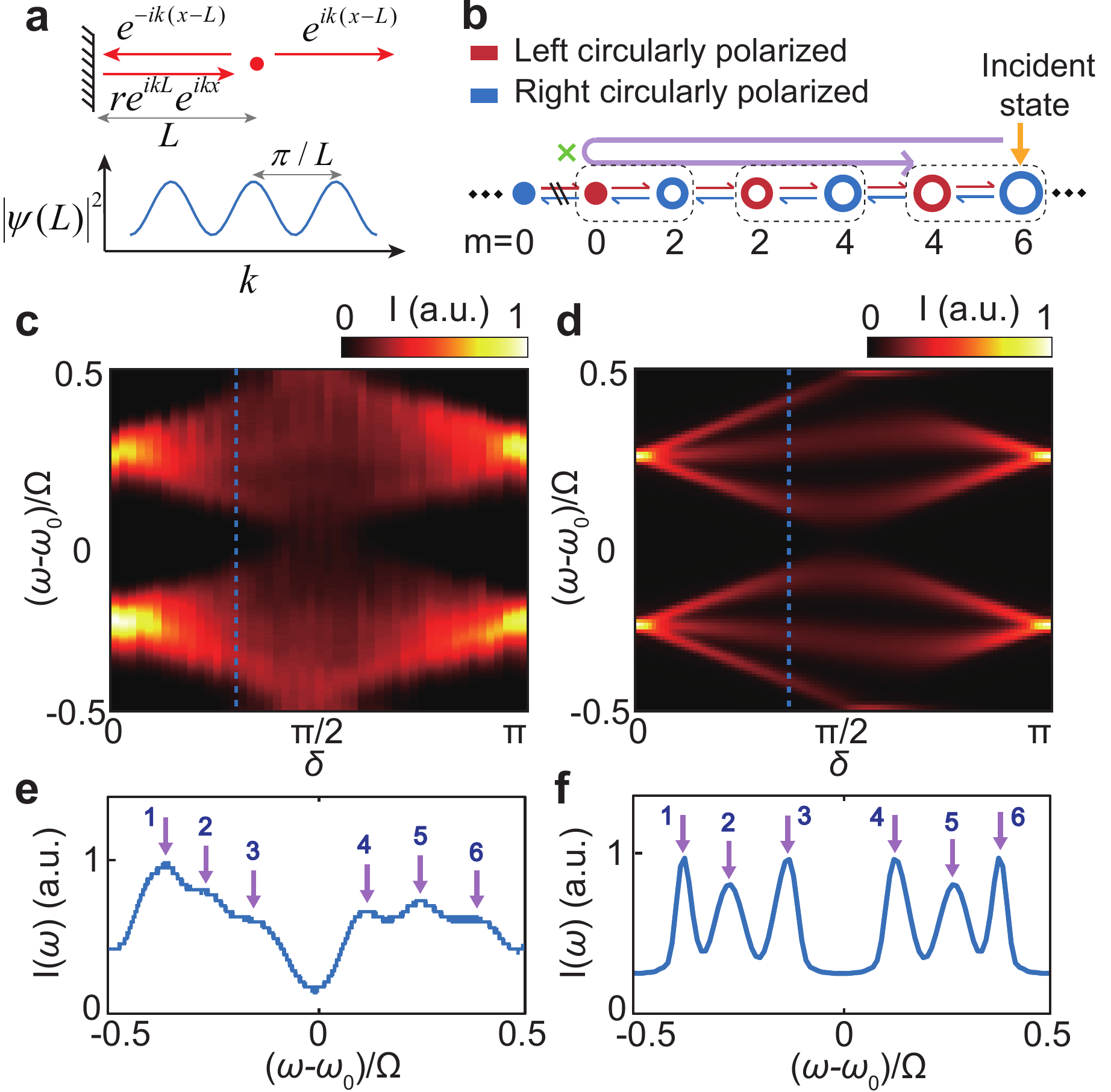}
\caption{\textbf{a}. Top panel: An emitter near the surface. Bottom panel: Oscillation of energy distribution $|\Psi(L)|^2$ caused by the interference.
\textbf{b}. Schematic of the lattice dynamics when exciting the left circularly polarized OAM mode with topological charge $m=6$. Photons travel along the purple arrow and the green cross indicates that the sharp boundary causes the reflection of the left-transmitted photons. \textbf{c, d}. The experimental (\textbf{c}) and numerically simulated (\textbf{d}) transmission intensity spectra. \textbf{e, f}. The experimental (\textbf{e}) and numerically simulated (\textbf{f}) transmission intensity spectrum when $\delta=0.35\pi$, corresponding to the dotted lines in \textbf{c} and \textbf{d}. $\omega-\omega_{0}$ is the frequency detuning, and $\Omega=375$ MHz represents the free spectral range (FSR).
}
\label{reflect}
\end{figure}

\textit{Spectrum discretization for the interference--}
 
As for an emitter close to a surface, the density of states will be modified because of the interference between the forward propagating and
the reflected wave~\cite{barnes1998fluorescence}. According to Bloch's theorem, the modes propagating on a periodic lattice can be approximated as a plane wave $e^{ikx}$, where $k$ is the momentum and $x$ is the axis. Here we consider a light source that will release two Bloch waves traveling in opposite directions $e^{\pm ik(x-L)}$, where $L$ represents the distance to the surface (Fig. \ref{reflect}a, top panel). Waves traveling to the left will be reflected and result in the state
\begin{equation}
\Psi(x)=e^{-ik(x-L)}+e^{ik(x-L)}+re^{ikL}e^{ikx},
\end{equation}
where $r\approx-1$ is the reflection coefficient. The interference between the leftward and the reflected waves causes an oscillation of the intensity distribution $|\Psi(x)|^2$. For example, the wave intensity at the position of emitter $|\Psi(L)|^2=5-4\cos{2kL}$ along with momentum $k$ is shown in Fig. \ref{reflect}a (bottom panel), where the oscillating period is determined by $\pi/L$. 

This effect will create spectrum discretization in our OAM lattice system. We pump the cavity with a high-order OAM mode and excite the right semi-infinite lattice site $n=\circlearrowright,6$ as shown Fig. \ref{reflect}b. The experimental and simulated transmitted intensity spectra $I(\omega)$ with different $\delta$ are shown in Fig. \ref{reflect}c and \ref{reflect}d, respectively. 
Since the left-transmitted mode and reflected mode interference in 3 unit cells (corresponding to $L=3$) between the boundary ($n=\circlearrowleft,0$) and the exciting sites ($n=\circlearrowright,6$), six discrete bands of the bulk are obtained in the transmitted spectra (see SM \Rom{4} for more simulation results). 
To clarify the distinct peaks for the transmitted spectra, we illustrate the contour of the transmitted intensity $I(\omega)$ with $\delta=0.35\pi$ as shown in Fig. \ref{reflect}e (experiment) and \ref{reflect}f (simulation), which show six discrete peaks in the bulk clearly. 

It's worth noting that the experimental transmitted peak has a wider peak width than the simulation results, which is more indistinguishable due to the loss of the cavity.

\textit{Conclusion---} We have experimentally demonstrated the sharp boundary along a synthetic OAM lattice and explored some topological boundary effects in a single optical cavity. We further explore the behavior of the edge state under perturbations, and we find that the spectrum discretization observed in our OAM lattice system can be well explained by the theory of emitters near the surface.   

Different from the previous finite lattice formed in real space~\cite{bandres2018topological,stutzer2018photonic} or synthetic space~\cite{kanungo2022realizing,dutt2022creating,lustig2019photonic}, our work provides an effective method to form  approximate semi-finite lattices. Notably, the properties of finite lattice and semi-finite lattice exhibit significant differences in non-Hermitian systems~\cite{yao2018edge, kawabata2019symmetry, gong2018topological}. 
Moreover, the degenerate cavity-based scheme makes it easier to support the couplings of multiple synthetic dimensions~\cite{yuan2019photonic} and cavities~\cite{luo2015quantum}, which provides opportunities to construct high dimensional topological insulators for exploring nontrivial topological boundary phenomena, such as the chiral edge states along the OAM lattice.
The long-range couplings along
the OAM lattice can also be introduced via a J-plate~\cite{devlin2017arbitrary}, which can perform arbitrary spin-to-orbital angular momentum conversion.
Drilling a hole in optical elements to distinguish different OAM modes can also build all-optical devices, such as all-optical memory~\cite{luo2017synthetic}. 
Furthermore, the degenerate cavity offers the possibility of coupling atoms~\cite{clark2020observation}, allowing for the exploration of effective many-particle systems by introducing photon-photon interactions via atomic mediums~\cite{PhysRevLett.79.1467, PhysRevLett.81.3611}. And our work provides a new topological photonic structure that may benefit topological devices such as topological lasers. \\

\begin{acknowledgments}
This work was supported by the Innovation Program for Quantum Science and Technology (Grants Nos.\,2021ZD0301400 and 2021ZD0301200), the National Natural Science Foundation of China (Grants Nos. 11821404, 92365205, U19A2075 and 61725504), Anhui Initiative in Quantum Information Technologies (AHY060300), the Fundamental Research Funds for the Central Universities (Grant No. WK5290000003). This work was partially carried
out at the USTC Center for Micro and Nanoscale Research and Fabrication. 
\end{acknowledgments}

\normalem
\bibliographystyle{jabbrv_unsrt}
\bibliography{edgebib}

\begin{thebibliography}{10}

\bibitem{mancini2015observation}
M.~Mancini, G.~Pagano, G.~Cappellini, L.~Livi, M.~Rider, J.~Catani, C.~Sias,
  P.~Zoller, M.~Inguscio, M.~Dalmonte, et~al.
\newblock Observation of chiral edge states with neutral fermions in synthetic
  hall ribbons.
\newblock {\em\JournalTitle{Science}}, 349(6255):1510--1513, 2015.

\bibitem{boada2012quantum}
O.~Boada, A.~Celi, J.~Latorre, and M.~Lewenstein.
\newblock Quantum simulation of an extra dimension.
\newblock {\em\JournalTitle{Physical Review Letters}}, 108(13):133001, 2012.

\bibitem{stuhl2015visualizing}
B.~Stuhl, H.-I. Lu, L.~Aycock, D.~Genkina, and I.~Spielman.
\newblock Visualizing edge states with an atomic bose gas in the quantum hall
  regime.
\newblock {\em\JournalTitle{Science}}, 349(6255):1514--1518, 2015.

\bibitem{yuan2016photonic}
L.~Yuan, Y.~Shi, and S.~Fan.
\newblock Photonic gauge potential in a system with a synthetic frequency
  dimension.
\newblock {\em\JournalTitle{Optics letters}}, 41(4):741--744, 2016.

\bibitem{dutt2019experimental}
A.~Dutt, M.~Minkov, Q.~Lin, L.~Yuan, D.~A. Miller, and S.~Fan.
\newblock Experimental band structure spectroscopy along a synthetic dimension.
\newblock {\em\JournalTitle{Nature Communications}}, 10(1):3122, 2019.

\bibitem{dutt2020single}
A.~Dutt, Q.~Lin, L.~Yuan, M.~Minkov, M.~Xiao, and S.~Fan.
\newblock A single photonic cavity with two independent physical synthetic
  dimensions.
\newblock {\em\JournalTitle{Science}}, 367(6473):59--64, 2020.

\bibitem{yang2022topological}
M.~Yang, H.-Q. Zhang, Y.-W. Liao, Z.-H. Liu, Z.-W. Zhou, X.-X. Zhou, J.-S. Xu,
  Y.-J. Han, C.-F. Li, and G.-C. Guo.
\newblock Topological band structure via twisted photons in a degenerate
  cavity.
\newblock {\em\JournalTitle{Nature Communications}}, 13(1):2040, 2022.

\bibitem{yang2023realization}
M.~Yang, H.-Q. Zhang, Y.-W. Liao, Z.-H. Liu, Z.-W. Zhou, X.-X. Zhou, J.-S. Xu,
  Y.-J. Han, C.-F. Li, and G.-C. Guo.
\newblock Realization of exceptional points along a synthetic orbital angular
  momentum dimension.
\newblock {\em\JournalTitle{Science Advances}}, 9(4):eabp8943, 2023.

\bibitem{chalabi2019synthetic}
H.~Chalabi, S.~Barik, S.~Mittal, T.~E. Murphy, M.~Hafezi, and E.~Waks.
\newblock Synthetic gauge field for two-dimensional time-multiplexed quantum
  random walks.
\newblock {\em\JournalTitle{Physical Review Letters}}, 123(15):150503, 2019.

\bibitem{weidemann2022topological}
S.~Weidemann, M.~Kremer, S.~Longhi, and A.~Szameit.
\newblock Topological triple phase transition in non-hermitian floquet
  quasicrystals.
\newblock {\em\JournalTitle{Nature}}, 601(7893):354--359, 2022.

\bibitem{regensburger2012parity}
A.~Regensburger, C.~Bersch, M.-A. Miri, G.~Onishchukov, D.~N. Christodoulides,
  and U.~Peschel.
\newblock Parity--time synthetic photonic lattices.
\newblock {\em\JournalTitle{Nature}}, 488(7410):167--171, 2012.

\bibitem{leefmans2022topological}
C.~Leefmans, A.~Dutt, J.~Williams, L.~Yuan, M.~Parto, F.~Nori, S.~Fan, and
  A.~Marandi.
\newblock Topological dissipation in a time-multiplexed photonic resonator
  network.
\newblock {\em\JournalTitle{Nature Physics}}, 18(4):442--449, 2022.

\bibitem{lustig2019photonic}
E.~Lustig, S.~Weimann, Y.~Plotnik, Y.~Lumer, M.~A. Bandres, A.~Szameit, and
  M.~Segev.
\newblock Photonic topological insulator in synthetic dimensions.
\newblock {\em\JournalTitle{Nature}}, 567(7748):356--360, 2019.

\bibitem{ozawa2019topological}
T.~Ozawa and H.~M. Price.
\newblock Topological quantum matter in synthetic dimensions.
\newblock {\em\JournalTitle{Nature Reviews Physics}}, 1(5):349--357, 2019.

\bibitem{yuan2018synthetic}
L.~Yuan, Q.~Lin, M.~Xiao, and S.~Fan.
\newblock Synthetic dimension in photonics.
\newblock {\em\JournalTitle{Optica}}, 5(11):1396--1405, 2018.

\bibitem{yang2022simulating}
M.~Yang, J.-S. Xu, C.-F. Li, and G.-C. Guo.
\newblock Simulating topological materials with photonic synthetic dimensions
  in cavities.
\newblock {\em\JournalTitle{Quantum Frontiers}}, 1(1):10, 2022.

\bibitem{luo2015quantum}
X.-W. Luo, X.~Zhou, C.-F. Li, J.-S. Xu, G.-C. Guo, and Z.-W. Zhou.
\newblock Quantum simulation of 2d topological physics in a 1d array of optical
  cavities.
\newblock {\em\JournalTitle{Nature Communications}}, 6(1):7704, 2015.

\bibitem{celi2014synthetic}
A.~Celi, P.~Massignan, J.~Ruseckas, N.~Goldman, I.~B. Spielman,
  G.~Juzeli{\=u}nas, and M.~Lewenstein.
\newblock Synthetic gauge fields in synthetic dimensions.
\newblock {\em\JournalTitle{Physical Review Letters}}, 112(4):043001, 2014.

\bibitem{sun2017weyl}
B.~Y. Sun, X.~W. Luo, M.~Gong, G.~C. Guo, and Z.~W. Zhou.
\newblock Weyl semimetal phases and implementation in degenerate optical
  cavities.
\newblock {\em\JournalTitle{Physical Review A}}, 96(1):013857, 2017.

\bibitem{luo2017synthetic}
X.-W. Luo, X.~Zhou, J.-S. Xu, C.-F. Li, G.-C. Guo, C.~Zhang, and Z.-W. Zhou.
\newblock Synthetic-lattice enabled all-optical devices based on orbital
  angular momentum of light.
\newblock {\em\JournalTitle{Nature Communications}}, 8(1):16097, 2017.

\bibitem{luo2018topological}
X.-W. Luo, C.~Zhang, G.-C. Guo, and Z.-W. Zhou.
\newblock Topological photonic orbital-angular-momentum switch.
\newblock {\em\JournalTitle{Physical Review A}}, 97(4):043841, 2018.

\bibitem{cardano2017detection}
F.~Cardano, A.~D’Errico, A.~Dauphin, M.~Maffei, B.~Piccirillo, C.~de~Lisio,
  G.~De~Filippis, V.~Cataudella, E.~Santamato, L.~Marrucci, et~al.
\newblock Detection of zak phases and topological invariants in a chiral
  quantum walk of twisted photons.
\newblock {\em\JournalTitle{Nature Communications}}, 8(1):15516, 2017.

\bibitem{wang2023proximity}
H.~Wang, K.~Murata, W.~Xie, J.~Li, J.~Zhang, K.~L. Wang, W.~Zhao, and T.~Nie.
\newblock Proximity-induced magnetic order in topological insulator on
  ferromagnetic semiconductor.
\newblock {\em\JournalTitle{Sci. China Inf. Sci.}}, 66(12):222403, 2023.

\bibitem{hatsugai1993chern}
Y.~Hatsugai.
\newblock Chern number and edge states in the integer quantum hall effect.
\newblock {\em\JournalTitle{Physical Review Letters}}, 71(22):3697, 1993.

\bibitem{kane2005z}
C.~L. Kane and E.~J. Mele.
\newblock Z$_{2}$ topological order and the quantum spin hall effect.
\newblock {\em\JournalTitle{Physical Review Letters}}, 95(14):146802, 2005.

\bibitem{thouless1982quantized}
D.~J. Thouless, M.~Kohmoto, M.~P. Nightingale, and M.~den Nijs.
\newblock Quantized hall conductance in a two-dimensional periodic potential.
\newblock {\em\JournalTitle{Physical Review Letters}}, 49(6):405, 1982.

\bibitem{dai2022topologically}
T.~Dai, Y.~Ao, J.~Bao, J.~Mao, Y.~Chi, Z.~Fu, Y.~You, X.~Chen, C.~Zhai,
  B.~Tang, et~al.
\newblock Topologically protected quantum entanglement emitters.
\newblock {\em\JournalTitle{Nature Photonics}}, 16(3):248--257, 2022.

\bibitem{kanungo2022realizing}
S.~Kanungo, J.~Whalen, Y.~Lu, M.~Yuan, S.~Dasgupta, F.~Dunning, K.~Hazzard, and
  T.~Killian.
\newblock Realizing topological edge states with rydberg-atom synthetic
  dimensions.
\newblock {\em\JournalTitle{Nature Communications}}, 13(1):972, 2022.

\bibitem{dutt2022creating}
A.~Dutt, L.~Yuan, K.~Y. Yang, K.~Wang, S.~Buddhiraju, J.~Vu{\v{c}}kovi{\'c},
  and S.~Fan.
\newblock Creating boundaries along a synthetic frequency dimension.
\newblock {\em\JournalTitle{Nature Communications}}, 13(1):3377, 2022.

\bibitem{zhou2017dynamically}
X.-F. Zhou, X.-W. Luo, S.~Wang, G.-C. Guo, X.~Zhou, H.~Pu, and Z.-W. Zhou.
\newblock Dynamically manipulating topological physics and edge modes in a
  single degenerate optical cavity.
\newblock {\em\JournalTitle{Physical Review Letters}}, 118(8):083603, 2017.

\bibitem{groning2018engineering}
O.~Gr{\"o}ning, S.~Wang, X.~Yao, C.~A. Pignedoli, G.~Borin~Barin, C.~Daniels,
  A.~Cupo, V.~Meunier, X.~Feng, A.~Narita, et~al.
\newblock Engineering of robust topological quantum phases in graphene
  nanoribbons.
\newblock {\em\JournalTitle{Nature}}, 560(7717):209--213, 2018.

\bibitem{le2020topological}
N.~H. Le, A.~J. Fisher, N.~J. Curson, and E.~Ginossar.
\newblock Topological phases of a dimerized fermi--hubbard model for
  semiconductor nano-lattices.
\newblock {\em\JournalTitle{npj Quantum Information}}, 6(1):24, 2020.

\bibitem{cheng2017degenerate}
Z.-D. Cheng, Z.-D. Liu, X.-W. Luo, Z.-W. Zhou, J.~Wang, Q.~Li, Y.-T. Wang,
  J.-S. Tang, J.-S. Xu, C.-F. Li, et~al.
\newblock Degenerate cavity supporting more than 31 laguerre--gaussian modes.
\newblock {\em\JournalTitle{Optics letters}}, 42(10):2042--2045, 2017.

\bibitem{cheng2018experimental}
Z.-D. Cheng, Q.~Li, Z.-H. Liu, F.-F. Yan, S.~Yu, J.-S. Tang, Z.-W. Zhou, J.-S.
  Xu, C.-F. Li, and G.-C. Guo.
\newblock Experimental implementation of a degenerate optical resonator
  supporting more than 46 laguerre-gaussian modes.
\newblock {\em\JournalTitle{Applied Physics Letters}}, 112(20), 2018.

\bibitem{Cheng:19}
Z.-D. Cheng, Z.-H. Liu, Q.~Li, Z.-W. Zhou, J.-S. Xu, C.-F. Li, and G.-C. Guo.
\newblock Flexible degenerate cavity with ellipsoidal mirrors.
\newblock {\em\JournalTitle{Opt. Lett.}}, 44(21):5254--5257, 2019.

\bibitem{barnes1998fluorescence}
W.~Barnes.
\newblock Fluorescence near interfaces: the role of photonic mode density.
\newblock {\em\JournalTitle{journal of modern optics}}, 45(4):661--699, 1998.

\bibitem{bandres2018topological}
M.~A. Bandres, S.~Wittek, G.~Harari, M.~Parto, J.~Ren, M.~Segev, D.~N.
  Christodoulides, and M.~Khajavikhan.
\newblock Topological insulator laser: Experiments.
\newblock {\em\JournalTitle{Science}}, 359(6381):eaar4005, 2018.

\bibitem{stutzer2018photonic}
S.~St{\"u}tzer, Y.~Plotnik, Y.~Lumer, P.~Titum, N.~H. Lindner, M.~Segev, M.~C.
  Rechtsman, and A.~Szameit.
\newblock Photonic topological anderson insulators.
\newblock {\em\JournalTitle{Nature}}, 560(7719):461--465, 2018.

\bibitem{yao2018edge}
S.~Yao and Z.~Wang.
\newblock Edge states and topological invariants of non-hermitian systems.
\newblock {\em\JournalTitle{Physical Review Letters}}, 121(8):086803, 2018.

\bibitem{kawabata2019symmetry}
K.~Kawabata, K.~Shiozaki, M.~Ueda, and M.~Sato.
\newblock Symmetry and topology in non-hermitian physics.
\newblock {\em\JournalTitle{Physical Review X}}, 9(4):041015, 2019.

\bibitem{gong2018topological}
Z.~Gong, Y.~Ashida, K.~Kawabata, K.~Takasan, S.~Higashikawa, and M.~Ueda.
\newblock Topological phases of non-hermitian systems.
\newblock {\em\JournalTitle{Physical Review X}}, 8(3):031079, 2018.

\bibitem{yuan2019photonic}
L.~Yuan, Q.~Lin, A.~Zhang, M.~Xiao, X.~Chen, and S.~Fan.
\newblock Photonic gauge potential in one cavity with synthetic frequency and
  orbital angular momentum dimensions.
\newblock {\em\JournalTitle{Physical Review Letters}}, 122(8):083903, 2019.

\bibitem{devlin2017arbitrary}
R.~C. Devlin, A.~Ambrosio, N.~A. Rubin, J.~B. Mueller, and F.~Capasso.
\newblock Arbitrary spin-to--orbital angular momentum conversion of light.
\newblock {\em\JournalTitle{Science}}, 358(6365):896--901, 2017.

\bibitem{clark2020observation}
L.~W. Clark, N.~Schine, C.~Baum, N.~Jia, and J.~Simon.
\newblock Observation of laughlin states made of light.
\newblock {\em\JournalTitle{Nature}}, 582(7810):41--45, 2020.

\bibitem{PhysRevLett.79.1467}
A.~Imamo\ifmmode~\bar{g}\else \={g}\fi{}lu, H.~Schmidt, G.~Woods, and
  M.~Deutsch.
\newblock Strongly interacting photons in a nonlinear cavity.
\newblock {\em\JournalTitle{Phys. Rev. Lett.}}, 79:1467--1470, Aug 1997.

\bibitem{PhysRevLett.81.3611}
S.~E. Harris and Y.~Yamamoto.
\newblock Photon switching by quantum interference.
\newblock {\em\JournalTitle{Phys. Rev. Lett.}}, 81:3611--3614, Oct 1998.

\end{thebibliography}
\newpage

\appendix{Supplementary materials}

\onecolumngrid
\setcounter{equation}{0}
\setcounter{figure}{0}
\renewcommand{\theequation}{S\arabic{equation}}
\renewcommand{\thefigure}{S\arabic{figure}}

\maketitle	
\tableofcontents

\section{Forming a Floquet SSH OAM lattice in a cavity}
Here we consider a closed standing wave degenerate cavity, which contains multiple orbital angular momentum (OAM) modes. There are two optical elements in the cavity, denoted as a Q-plate and a wave plate (WP). The Q-plate operator on the polarized OAM states can be described as
\begin{equation}
\begin{aligned}
J_{Q}=\sum_{m}\cos(\delta/2)(a_{\circlearrowleft,m}^{\dag}a_{\circlearrowleft,m}+a_{\circlearrowright,m}^{\dag}a_{\circlearrowright,m})+i\sin(\delta/2)(a_{\circlearrowright,m+2q}^{\dag}a_{\circlearrowleft,m}+\mathrm{h.c.}),
\end{aligned}\label{Qp}
\end{equation}
where $\delta$ is the electrically controlled parameter, and $\sin{(\delta/2)}$ is the coupling strength between adjacent OAM modes. $a_{\circlearrowleft(\circlearrowright),m}$ and $a_{\circlearrowleft(\circlearrowright),m}^{\dag}$
are annihilation and creation operators of left/right circular polarizations. Here $q=1$ in the experiment, which means only OAM modes with even topological charge will be coupled by the Q-plate with even order OAM modes incidence.

Similarly, the operator of the wave plate introducing hopping between two polarized states with the same OAM is
\begin{equation}
\begin{aligned}
J_{W}=\sum_{m}\cos(\eta/2)(a_{\circlearrowleft,m}^{\dag}a_{\circlearrowleft,m}+a_{\circlearrowright,m}^{\dag}a_{\circlearrowright,m})+i\sin(\eta/2)(e^{-i\xi}a_{\circlearrowright,m}^{\dag}a_{\circlearrowleft,m}+e^{i\xi}a_{\circlearrowleft,m}^{\dag}a_{\circlearrowright,m}),
\end{aligned}
\end{equation}
where $\eta$ denotes the phase delay between ordinary and extraordinary photons and is determined by the thickness of the quartz plate. $\xi$ is the angle of the wave plate optical axis. As we drill a pinhole on the WP where only the OAM mode with $m=0$ can pass through, the operator $J_W^{hole}$ of the WP is expressed as
\begin{equation}
\begin{aligned}
J_{W}^{hole}=&\sum_{m\neq0}\cos(\eta/2)(a_{\circlearrowleft,m}^{\dag}a_{\circlearrowleft,m}+a_{\circlearrowright,m}^{\dag}a_{\circlearrowright,m})+i\sin(\eta/2)(e^{-i\xi}a_{\circlearrowright,m}^{\dag}a_{\circlearrowleft,m}+e^{i\xi}a_{\circlearrowleft,m}^{\dag}a_{\circlearrowright,m})\\
&+a_{\circlearrowleft,0}^{\dag}a_{\circlearrowleft,0}+a_{\circlearrowright,0}^{\dag}a_{\circlearrowright,0}.
\end{aligned}
\end{equation}

In a round trip in the cavity, the light field double passes through the Q-plate and the WP, and the evolution of the polarized OAM states can be written as 
\begin{equation}
\begin{aligned}
U=J_{Q}J^{hole}_{W}J^{hole}_{W}J_{Q}.
\end{aligned}
\end{equation}
For the periodic-driven nature of the system, we can define an effective Hamiltonian satisfying $\hat{H}_{\rm eff}=-i\ln{U}$. To analyze the eigenvalues and states of the corresponding evolution, we take the wave function of the polarized OAM state as
\begin{equation}
 |\psi\rangle = (\cdots, \psi_{-m,\circlearrowright},\psi_{-m,\circlearrowleft},\cdots,\psi_{-2,\circlearrowright},\psi_{-2,\circlearrowleft},\psi_{0,\circlearrowright},\psi_{0,\circlearrowleft}, \psi_{2,\circlearrowright},\psi_{2,\circlearrowleft},\cdots,\psi_{m,\circlearrowright},\psi_{m,\circlearrowleft},\cdots).
\end{equation}
The operators of the Q-plate and WP can be written as large matrices, given by
\begin{equation}
J_{Q}=
\left(
\begin{array}{cccc}
\cdots&0&0&0\\
i\sin(\frac{\delta}{2}) & \cos(\frac{\delta}{2})  & 0 & 0 \\
0 & 0 & \cos(\frac{\delta}{2}) & i\sin(\frac{\delta}{2})\\
0&0&0&\cdots\\
\end{array}
\right),
\end{equation}
and 
\begin{equation} 
J_{W}^{\rm hole}=\left(
\begin{array}{cccccccc}
\cdots&0&0&0&0&0&0&0 \\
0&\cos(\frac{\eta}{2})  & \sin(\frac{\eta}{2}) & 0 & 0 & 0 & 0 & 0 \\
0&-\sin(\frac{\eta}{2}) & \cos(\frac{\eta}{2}) &0&0&0&0&0 \\
0&0&0&1&0&0&0&0 \\
0&0&0&0&1&0&0&0 \\
0&0&0&0&0&\cos(\frac{\eta}{2})  & \sin(\frac{\eta}{2}) &0   \\
0&0&0&0&0&-\sin(\frac{\eta}{2}) & \cos(\frac{\eta}{2}) &0 \\
0&0&0&0&0&0&0&\cdots\\
\end{array}
\right).
\label{WPk}
\end{equation}
By solving the stationary Schrodinger equation 
\begin{equation}
U|\psi\rangle =e^{iH_{\rm eff}}|\psi\rangle=e^{iE}|\psi\rangle, 
\end{equation}
we can get the eigen-energy $E$ and eigen-states $|\psi\rangle$. As an example, the solved eigen-energy ($E$) along $\delta$ ($\delta\in(-\pi,\pi]$) with $\eta=\pi/2$ are shown in Fig. \ref{sim}a. We can clearly find the zeros-energy edge modes when $|\delta|<\pi/2$ and $\pm\pi$-energy edge modes when $|\delta|>\pi/2$. The solved eigenstates $|\psi\rangle$ as $\delta=9\pi/40$ (corresponding to the eigen-energy of pink points in Fig. \ref{sim}a) are shown in Fig. \ref{sim}b. The bulk states are evenly distributed (blue bars) along the OAM lattice while the zero energy edge modes (colored by orange bars) are mostly located at the $m=0$ site. 

\begin{figure*}[t]
\centering\includegraphics[width=15cm]{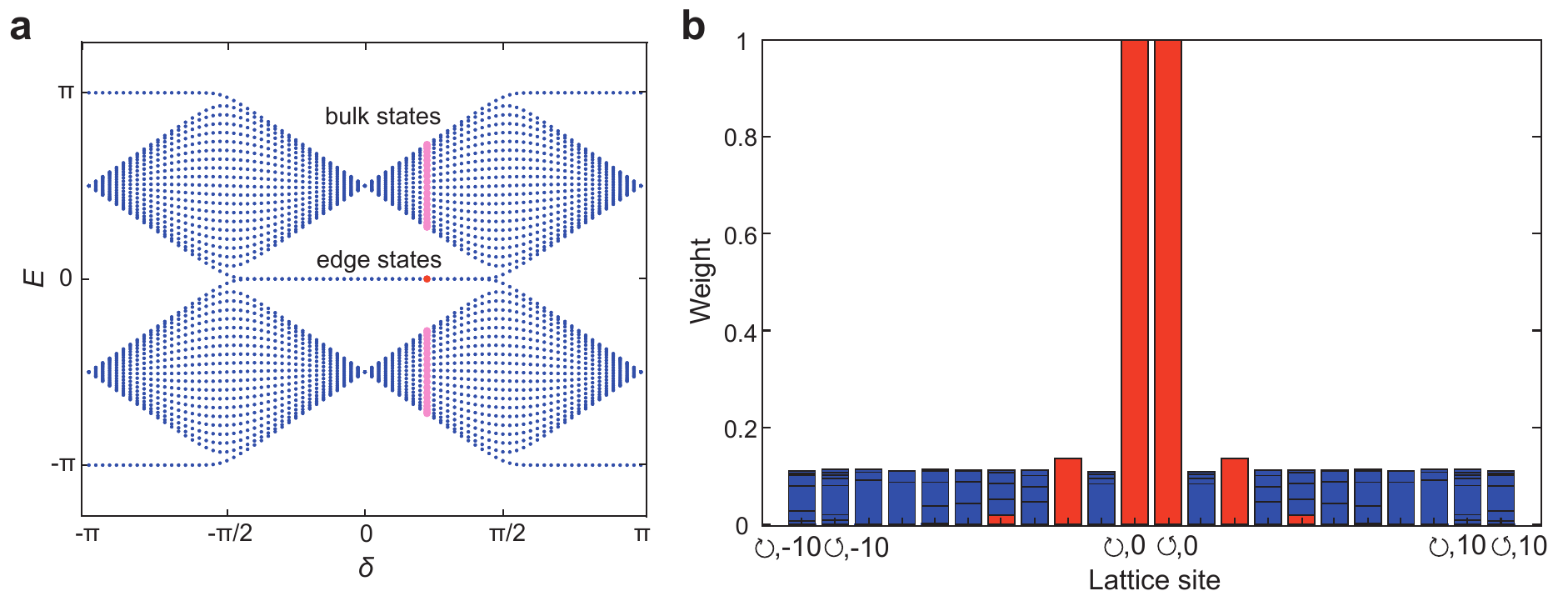}
\caption{\textbf{The calculated eigen-energies and eigen-states when $\eta=\pi/2$.} \textbf{a}. The eigenenergies along different $\delta$. \textbf{b}. The eigenstates distribution as $\delta=9\pi/40$, corresponding to the pink points in \textbf{a}. The orange bars correspond to the zero energy edge states while the blue bars correspond to bulk states. }
\label{sim}
\end{figure*}

\section{Tightly bounding approximation of the Hamiltonian}
On the other hand, as the coupling strength is small, we can write the analytic expression of the system Hamiltonian directly. As $\delta\ll1$, the operator of the Q-plate can be approximated to
\begin{equation}
\begin{aligned}
J_{Q}\approx 1+i\sum_{m}\frac{\delta}{2}(a_{\circlearrowright,m+2q}^{\dag}a_{\circlearrowleft,m}+\mathrm{h.c.})
=e^{i\sum_{m}\frac{\delta}{2}(a_{\circlearrowright,m+2q}^{\dag}a_{\circlearrowleft,m}+\mathrm{h.c.})}.
\end{aligned}\label{Qp}
\end{equation}
Similarly, as $\eta\ll1$, the operator of the WP can be approximated to
\begin{equation}
\begin{aligned}
J_{W}^{hole}&\approx 1+i\sum_{m\neq0}\frac{\eta}{2}(a_{\circlearrowright,m}^{\dag}a_{\circlearrowleft,m}-a_{\circlearrowleft,m}^{\dag}a_{\circlearrowright,m})
=e^{i\sum_{m\neq0}\frac{\eta}{2}(a_{\circlearrowright,m}^{\dag}a_{\circlearrowleft,m}-a_{\circlearrowleft,m}^{\dag}a_{\circlearrowright,m})},
\end{aligned}
\end{equation}
Thus, the evolution $U$ of the light field in a round trip can be written as
\begin{equation}
\begin{aligned}
U&=J_{Q}J^{hole}_{W}J^{hole}_{W}J_{Q}\\
&=e^{i\sum_{m}\delta(a_{\circlearrowright,m+2q}^{\dag}a_{\circlearrowleft,m}+\mathrm{h.c.})+i\sum_{m\neq0}\eta(a_{\circlearrowright,m}^{\dag}a_{\circlearrowleft,m}-a_{\circlearrowleft,m}^{\dag}a_{\circlearrowright,m})}
\end{aligned}
\end{equation}
The system Hamiltonian is 
\begin{equation}
\begin{aligned}
\hat{H}&=-i\ln{U}\\
&=\sum_{m}\delta(a_{\circlearrowright,m+2q}^{\dag}a_{\circlearrowleft,m}+\mathrm{h.c.})+\sum_{m\neq0}\eta(a_{\circlearrowright,m}^{\dag}a_{\circlearrowleft,m}-a_{\circlearrowleft,m}^{\dag}a_{\circlearrowright,m}).
\end{aligned}
\end{equation}

\section{Input and output relations of the cavity}
In the experiment of edge state in the main text, we measure the system by detecting the output signal of the degenerate cavity which is pumped by a continuous-wave laser. To study the characteristics of the transmission process we consider the interaction between the cavity modes and the outside optical modes. The total Hamiltonian has the form:
\begin{equation}
\hat{H}=\hat{H}_{sys}+\hat{H}_{b}+\hat{H}_{int},
\label{htotal}
\end{equation}
where $\hat{H}_{sys}$ is the Hamiltonian of the cavity system. $\hat{H}_b$ is the heat bath Hamiltonian which describes the outside field as:
\begin{equation}
\hat{H}_b=\sum_{n}\int_{-\infty}^{\infty} d\omega \hbar \omega b^{\dag}_n(\omega)b_n(\omega),
\end{equation}
where $b_n(\omega)$ are boson annihilation operators for the bath, label $n=(s,m)$ denote the polarization and the OAM number of the photon modes. $b_n(\omega)$ obey the commutation relation:
\begin{equation}
[b_n(\omega),b^{\dag}_{n'}(\omega')]=\delta(\omega-\omega')\delta_{nn'}.
\end{equation}
$\hat{H}_{int}$ in \ref{htotal} describes the coupling of the system to the bath. With rotating wave approximation and assuming the coupling constant $\kappa(\omega)=\sqrt{\gamma/2\pi}$ is independent of the frequency, $\hat{H}_{int}$ has the form:
\begin{equation}
\hat{H}_{int}=-i\hbar \sum_n \int_{-\infty}^{\infty}d\omega \sqrt{\frac{\gamma}{2\pi}}[a_n^{\dag}b_n(\omega)-b_n^{\dag}(\omega)a_n],
\end{equation}
where $a_n$ is the annihilation operator of the photon mode in the system. 

Defining in field operators:
\begin{equation}
    b_{in,n}(t)=\frac{1}{\sqrt{2\pi}}\int_{-\infty}^{\infty}d\omega e^{-i\omega t}b_{n,0}(\omega),
\end{equation}
with $b_{n,0}(\omega)$ the value of $b_{n}(\omega)$ at $t=0$, which also satisfy: $[b_{in,n}(t),b_{in,n^{'}}^{\dag}(t')]=\delta_{nn'}\delta(t-t')$. We can write down the Langevin equation:
\begin{equation}
    \frac{da_n(t)}{dt}=-i[a_n,\hat{H}_{sys}]-\frac{\gamma}{2}a_n(t)-\sqrt{\gamma}b_{in,n}(t).
\label{L_equation}
\end{equation}
We also have the input-output relation:
\begin{equation}
    b_{out,n}(t)-b_{in,n}(t)=\sqrt{\gamma}a_n(t),
\label{in-out}
\end{equation}
where $b_{out,n}(t)=\frac{1}{\sqrt{2\pi}}\int_{-\infty}^{\infty}d\omega e^{-i\omega (t-t_1)}b_{n,1}(\omega)$ are defined as the out field operators, and $b_{n,1}(\omega)$ is the boson annihilation operator at time $t_1>t$. 

Combine \ref{L_equation} and \ref{in-out}, transfer them to the frequency domain using Fourier transformation, and we get the solution:
\begin{equation}
    b_{out,n'}(\omega)=\sum_n(\delta_{nn'}-i\langle n'|\frac{\gamma}{\omega-\hat{H}_{sys}+\frac{i\gamma}{2}}|n\rangle)b_{in,n}(\omega).
\label{solution}
\end{equation}
$|n\rangle=a^{\dag}_n$ in \ref{solution} is a single photon mode. The first term on the right side represents the reflection and the second term is the transmission part. With a single mode input field $b_{in}(\omega)=b_{in,n}(\omega)$, the transmission coefficient takes the form: 
\begin{equation}
    T_{n'}=-i\langle n'|\frac{\gamma}{\omega-\hat{H}_{sys}+\frac{i\gamma}{2}}|n\rangle.
\end{equation}

\section{Simulations of the bands with interference}
\begin{figure}[tb]
	\begin{center}		
    \includegraphics[width=1 \columnwidth]{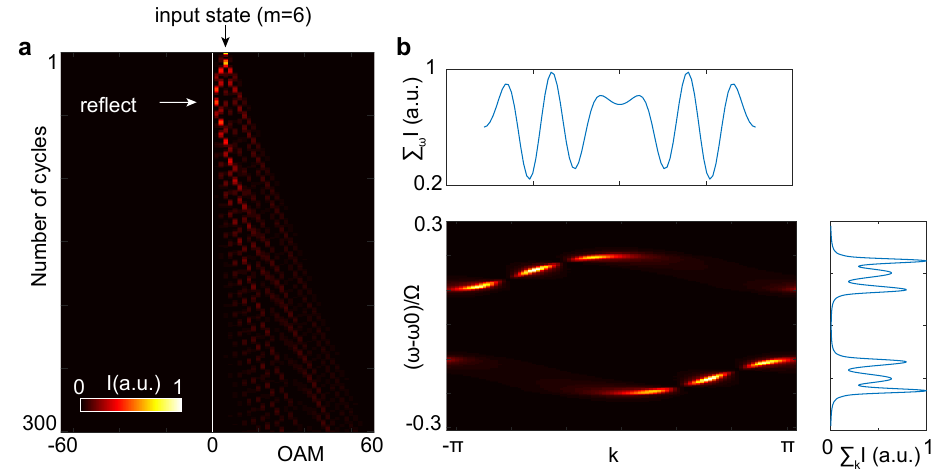}
    \caption{\textbf{a}. OAM distributions of the light circulating in the cavity. \textbf{b}. Normalized transmission intensity spectra vs momentum $k$. top (right) panel: total normalized transmission intensity spectra. }
		\label{sim}
	\end{center}
\end{figure}

To show the influence on the transmitted spectra of the interference in the OAM lattice, we numerically simulate the evolution of OAM modes in the cavity. 
As we input the OAM modes $|\phi_{in}\rangle$ with topological charge $m_{0}=6$ into the cavity (where the sites between excited point and the boundary form 3 unit cells corresponding to L=3, as shown in Fig. 4b), the OAM states start to evolve according to $\hat{U}^{n}|\phi_{in}\rangle$, where $n$ represents the number of times light circulates in the cavity. The OAM distribution at each cycle is shown in Fig. \ref{sim}a, and we can find that leftward waves along the OAM lattice meet the boundary (m=0) and reflect after three cycles. 
The output of the cavity is a coherent superposition of each cyclic state, which can be written as $|\phi_{out}\rangle=\kappa\sum_{n}t^{n-1}e^{-i2n\pi(\omega-\omega_{0})/\Omega}\hat{U}^{n}|\phi_{in}\rangle$, where $\kappa$ and $t$ are the coupling coefficient and reflection coefficient of the cavity mirror satisfying $|\kappa|^2+|t|^2=1$. $\omega-\omega_{0}$ is the detuning of the input frequency and $\Omega$ is the free spectral range (FSR). As we project the output states on the Bloch wave vector $|k\rangle=\sum_{m}e^{-ikm}|m\rangle$, we can get the normalized transmission intensity spectra $I(\omega,k)=\langle\phi_{out}|k\rangle\langle k|\phi_{out}\rangle$ along momentum $k$ (Fig. \ref{sim}b). We can find the spectrum for the leftward modes ($v_g=\partial\omega/\partial k<0$) split into three peaks for the interference between the leftward propagating and the reflected modes. Thus the total normalized transmission intensity $\sum_{\omega}I(\omega,k)$ along momentum $k$ has six peaks (Fig. \ref{sim}b, top), which is consistent with the prediction of the formula $|\Psi(L=3)|^2$ in Fig. 4a. In our experiment, we detect the total normalized transmission intensity spectra ($\sum_{k}I(\omega,k)$) containing all $k$ and we can also obtain the six peaks (Fig. \ref{sim}b, right).

\section{Details of the experimental setup to create and measure the edge states}
\begin{figure*}[t]
\centering\includegraphics[width=15cm]{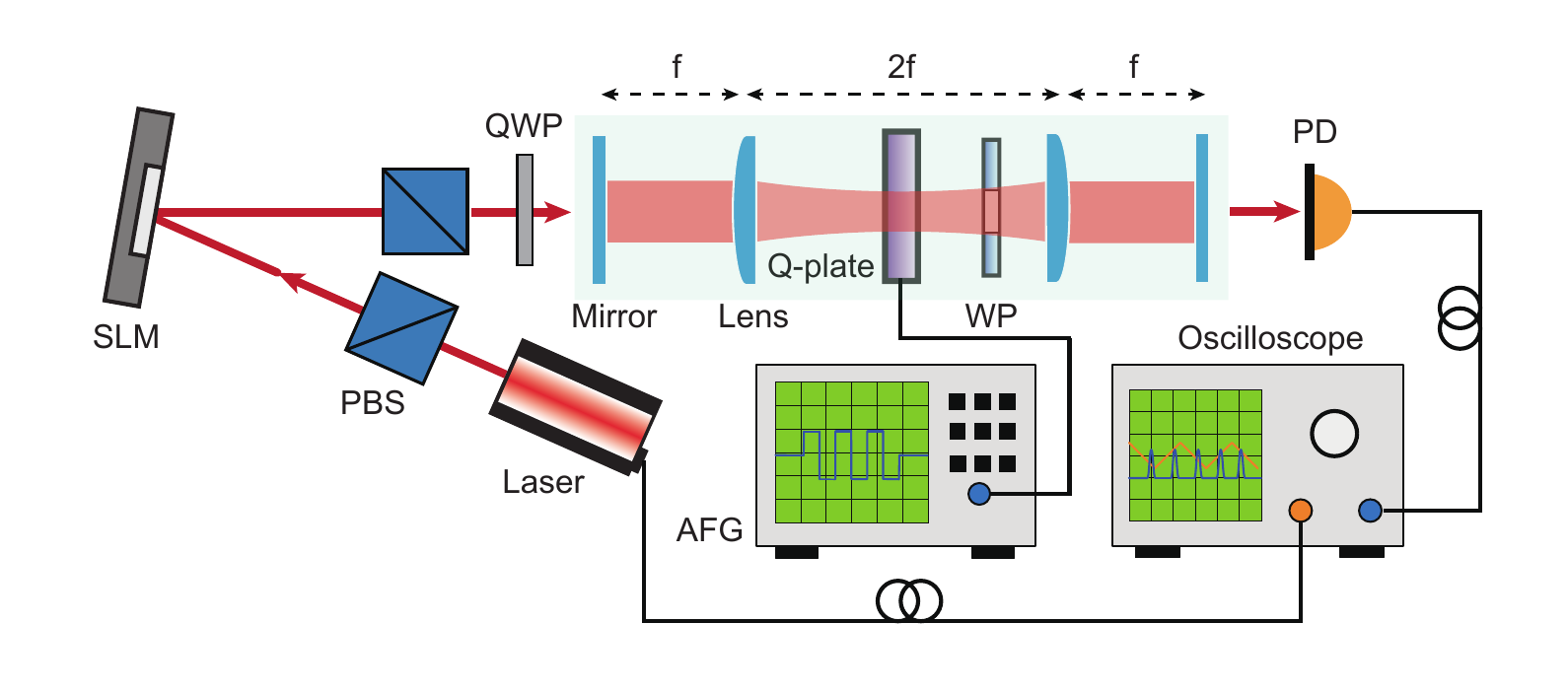}
\caption{The experimental setup. SLM: spatial light modulator; PBS: polarizing beamsplitter; AFG: arbitrary function generator; PD: photon detector; QWP: quarter-wave plate; WP: wave plate.}
\label{expset}
\end{figure*}

The experimental setup for edge state detecting is shown in Fig. \ref{expset}. We use a degenerate cavity to form the synthetic OAM dimension. The cavity transverse modes are Laguerre-Gaussian (LG) modes satisfying 
\begin{equation}
\begin{aligned}
E(r, \phi, z)=\frac{C_{m p}^{L G}}{w(z)}\left(\frac{r \sqrt{2}}{w(z)}\right)^{|m|} e^{\left(-\frac{r^2}{w^2(z)}\right)} L_p^{|m|}\left(\frac{2 r^2}{w^2(z)}\right) e^{-i k \frac{r^2}{2 R(z)}} e^{i(2p+|l|+1)\zeta(z)} e^{im\phi},
\end{aligned}
\end{equation}
where $p$ and $m$ are radial and angular indices. $w(z)=w_{0}\sqrt{1 +(z/z_{R})^2}$ represents beam size, where $w_{0}$ is waist size. $z_{R} $ is Rayleigh distance and $\zeta(z)=\arctan(z/z_{R})$ is Gouy phase shift. $L_p^{|m|}$ represents the Laguerre polynomial, and $R(z)$ represents the radius of curvature of the wavefront. The resonant frequency of the Laguerre-Gaussian modes in the cavity is:
\begin{equation}
\begin{aligned}\label{D}
2\pi\frac{\omega}{\Omega}-(2p+l+1)arccos(\frac{A+D}{2})=2n\pi,
\end{aligned}
\end{equation}
where $\omega$ is the light frequency and $\Omega$ is the free spectral range (FSR) of the cavity. $A$ and $D$ are diagonal elements of the transmission matrix of the cavity. A cavity that satisfies $(A+D)/2=1$ is defined as a degenerate cavity, where all spatial modes of the degenerate cavity share the same resonant frequency.
The cavity consists of two plane mirrors, two lenses, a Q-plate, and a wave plate. For the coupled-in mirror, the ratio between transmittance and reflectance is 5/95, and for the coupled-out mirror, it's 1/99. Two lenses in the cavity have the same focal length $f=10$ cm and form a 4f telescope (corresponding to $(A+D)/2=1$), which holds the cavity degeneracy to all the OAM modes. The Q-plate introduces the hopping between adjacent OAM modes, and the parameter $\delta$ is controlled by a square wave signal applied by an arbitrary function generator (AFG). The wave plate has a hole in the center with a radius $r=125$ $\upmu$m. To pump the system, we use a laser, which has a central wavelength of 880 nm, combined with a spatial light modulator (SLM) to generate light of different OAM modes. Holograms of vortex beam with different topological charges m are loaded on the SLM. The incident light to the SLM is polarized to the horizontal polarization by the polarizing beamsplitter (PBS). After modulating, the light passes through a 4f system (not illustrated in the figure). And in the middle of the two lenses, we place a pinhole to filter out unmodulated light. To measure the transmission spectrum of the cavity, we sweep the frequency of the laser from 0 to 500 MHz by a triangle wave signal, which is also used to trigger the oscilloscope. A photon detector (PD) connected to the oscilloscope detects the transmitted photons. 

\section{Fabrication of the pinhole on the wave plate}

\begin{figure}[hbt]
	\begin{center}
		\includegraphics[width=0.8 \columnwidth]{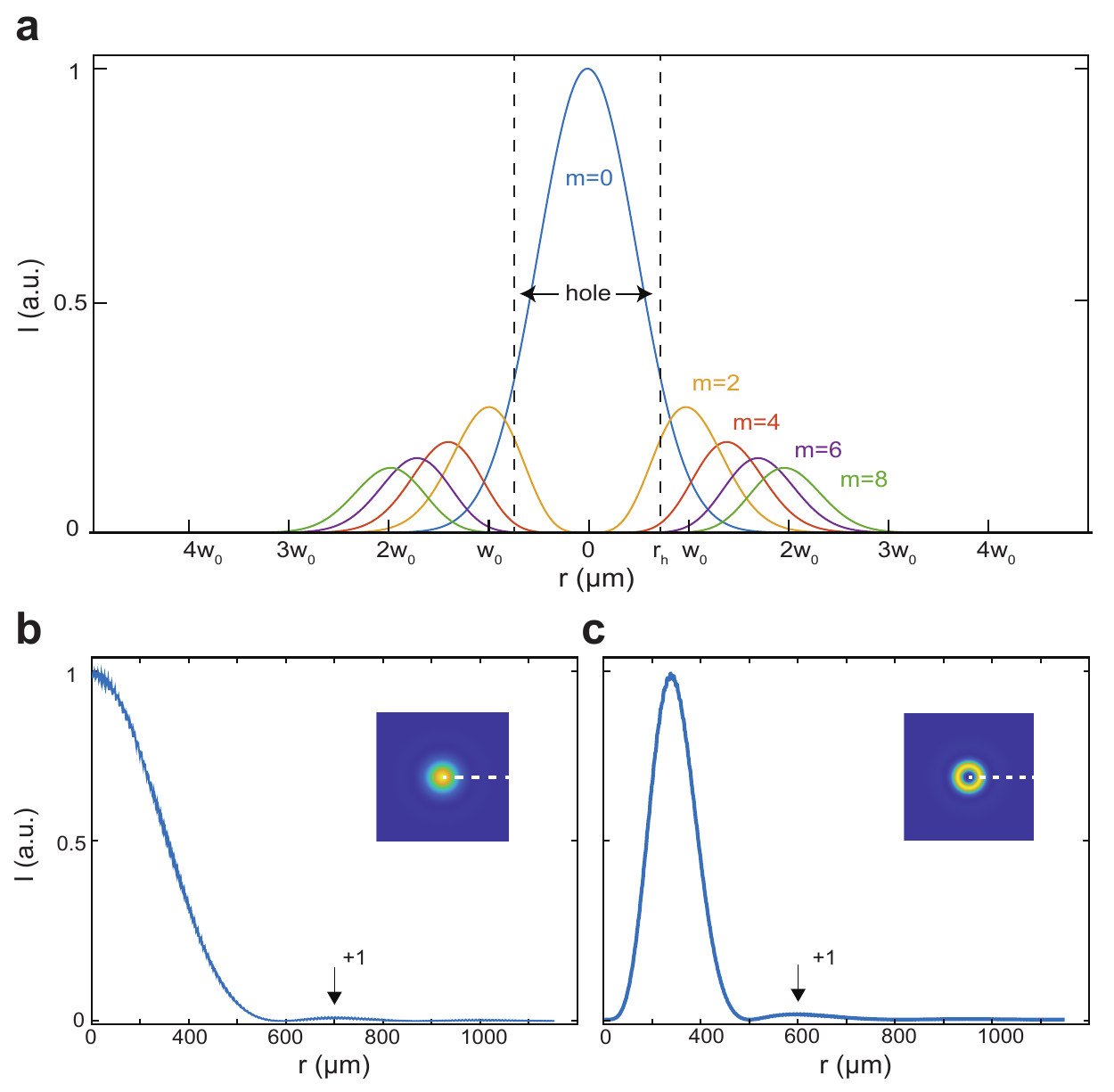}
		\caption{\textbf{a}. The intensity distributions of different cavity OAM modes. The dotted lines show the size of the pinhole. \textbf{b,c}. Inserts: the diffraction patterns of the OAM modes with different topological charges $m=0$ (b) and $m=2$ (c). Curves: Normalized intensity profiles along the white dotted line. }
		\label{edge}
	\end{center}
\end{figure}

The wave plate with a central circular hole is the core optical element to create the boundary along the synthetic OAM dimension, and we need to rightfully determine the size of the pinhole. 
The intensity profiles of different cavity transverse OAM modes near the waist are shown in Fig. \ref{edge}a, which quantifies the overlap integral between the different cavity transverse OAM modes and the spatial aperture in the waveplate. By carefully preparing the pinhole with a specific radius $r_{h}=0.75w_{0}=125$ $\upmu$m by using femtosecond laser processing technology in experiments, where $w_0$ is the waist of the incident LG beam, we can minimize the scattering of the small residual portion of light for $m\neq0$ modes. Here we simulate the diffraction of OAM modes with $m=0$ (Fig. \ref{edge}b) and $m=2$ (Fig. \ref{edge}c). Indeed, the residual portion of modes that interacts with the aperture results in a small +1 order diffraction ring (marked by black arrows), but the relative intensity is small ($I_{+1}/I_{all}<2\%$). Thus the extra losses or spurious modal cross-coupling terms can be almost ignored. 

Then we use a femtosecond laser to ablate a pinhole on a true zero-order three-fourths wave plate with a 75 $\upmu$m thickness. 

The wavelength of the femtosecond laser is 1030 nm. The pulse width and repetition frequency are 1000 fs and 19.9 kHz. The focus depth is set at 5 $\upmu$m, 35 $\upmu$m, and 70 $\upmu$m, respectively, and the laser is scanned ten times at each depth. The well-fabricated wave plate must be ultrasonically cleaned for 5 minutes to prevent the blockage of small holes.

\section{Detecting the edge state distribution via projective measurement}

\begin{figure}[hbt]
	\begin{center}
		\includegraphics[width=1 \columnwidth]{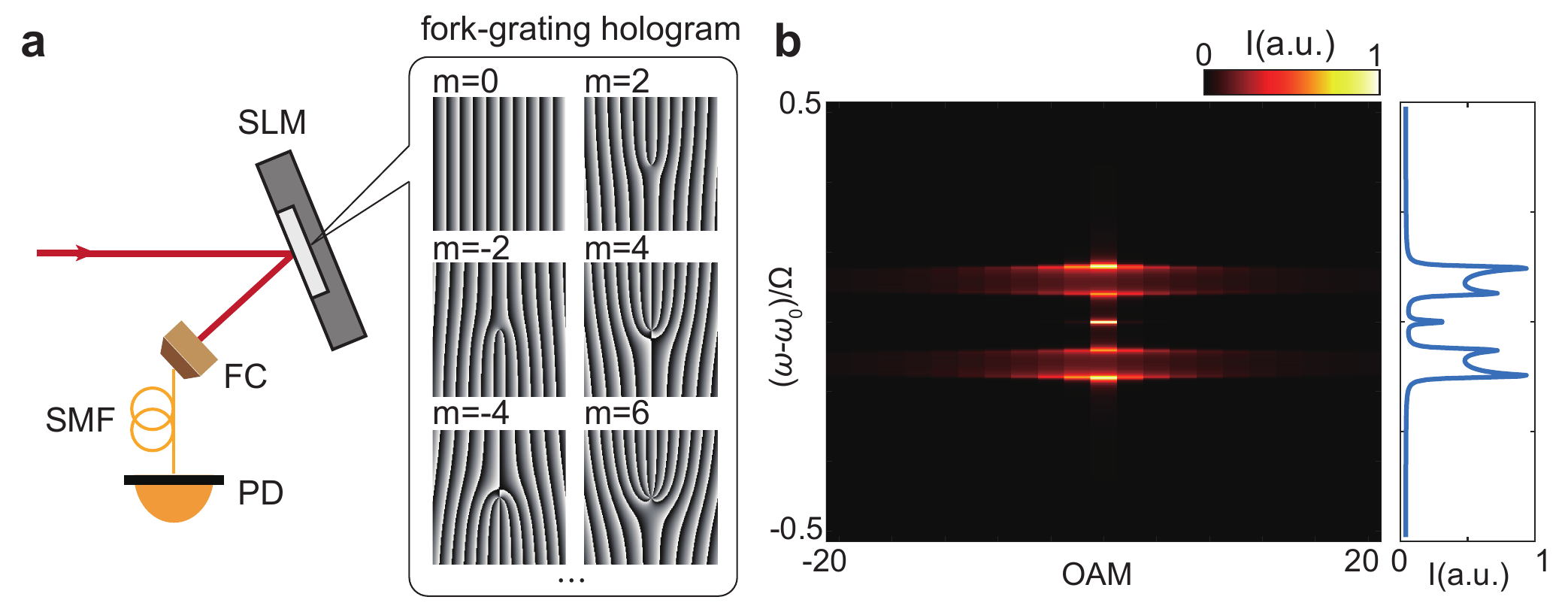}
		\caption{\textbf{a}. The experimental setup of projective measurement to measure edge state spatial distribution. SLM: spatial light modulator; PD: photon detector; FC: fiber coupler; SMF: single-mode fiber; \textbf{b}. Left: the simulated transmission intensity spectrum on different OAM bases with $\delta=0.1$ and $\eta=0.2$. Right: the simulated total transmission intensity spectrum. }
		\label{projection}
	\end{center}
\end{figure}

Our platform can reveal edge state spatial distribution by projection measurement (post-selecting method), which is implemented by the configuration in Fig. \ref{projection}a. A spatial light modulator (SLM) loaded with different fork-grating holograms converts the OAM mode to Gaussian mode, then a single-mode fiber collects the optical signal (only Gaussian mode can be coupled into single-mode fiber). By changing the topological charge $m$ of the fork-grating holograms, different OAM projection bases can be achieved. 
A simulated example is given in Fig. \ref{projection} with $\delta=0.1$ and $\eta=0.2$. The total intensity spectrum is shown in Fig. \ref{projection} (right), where the edge state is at the center of an energy gap. After projection measurement, we can obtain the transmission intensity spectra (Fig. \ref{projection}, left) on different OAM bases, where the edge mode of $E=0$ is mainly distributed as the OAM mode of $m=0$, which is consistent with the results (Fig. 1d) via the pre-selection method we used.

\end{document}